\documentclass[12pt,aps,prl,notitlepage,groupedaddress,tightenlines,onecolumn,amsmath,amssymb,floatfix,nofootinbib]{revtex4-1}
\pdfoutput=1 
\usepackage{amsmath}
\usepackage{amssymb}
\usepackage{graphicx}
\usepackage[caption=false]{subfig}
\let\oldFootnote\footnote
\newcommand\nextToken\relax
\renewcommand\footnote[1]{%
    \oldFootnote{#1}\futurelet\nextToken\isFootnote}
\newcommand\isFootnote{%
    \ifx\footnote\nextToken\textsuperscript{,}\fi}
\def\eq{\begin{equation}}
\def\en{\end{equation}}
\def\partialby#1{\frac{\partial\hfill}{\partial#1}}
\def\Reals{\mathbb{R}}

\def\SO{\mathrm{SO}}

\begin{document}
\title{Cosmology from the two-dimensional renormalization group 
acting as the Ricci flow}
\author{Daniel Friedan}
\email[]{dfriedan@gmail.com}
\homepage[]{\path{www.physics.rutgers.edu/~friedan/}}
\affiliation{New High Energy Theory Center and Department of Physics and Astronomy,
Rutgers, The State University of New Jersey,
Piscataway, New Jersey 08854-8019 U.S.A.}
\affiliation{Science Institute, The University of Iceland,
Reykjavik, Iceland
%\thanks{}
}
\date{September 3, 2019}
\begin{abstract}
The two-dimensional renormalization group acting as the Ricci flow
$\Lambda\partialby{\Lambda} g_{\mu\nu} = R_{\mu\nu}$ produces a 
specific 1+3
dimensional space-time metric which describes an expanding universe
that starts with a big bang $a \sim t^{\scriptscriptstyle 1/\sqrt3}$ then decelerates until
$z=0.2$ then accelerates until ending at $t_{\max}=1.6\,t_{H}$ with a big
blowup $a \sim (t_{\max}-t)^{\scriptscriptstyle -1/\sqrt3}$.  The only free 
parameters are
the overall time scale and the value of the present time
$t_{0}$.  These are fixed by the Hubble constant
$H_{0}=t_{H}^{-1}$ and the present deceleration parameter
$q(t_{0})$.  This crude calculation of cosmology omits all 
but the
gravitational field.  The only energy-momentum is purely gravitational dark
matter and energy.  
This is a  preliminary exploration towards
a specific, comprehensive, testable calculation of cosmology from a
fundamental theory in which physics is produced by a quantum version
of the two-dimensional renormalization group.
\end{abstract}
%\keywords{}
%
\maketitle

The renormalization group of the two-dimensional general nonlinear
model acts at leading order as the Ricci flow
\cite{Friedan1980a,Friedan1980c,Friedan:1980jmcosmo}
\eq
\Lambda\partialby{\Lambda} g_{\mu\nu} = R_{\mu\nu}
\en
The field of the nonlinear model takes its values in a Riemannian
manifold $M$.  The Riemannian metric $g_{\mu\nu}$ encodes the
couplings of the model.  The beta-function at leading order is the Ricci tensor 
$R_{\mu\nu}$.  The renormalization group drives $g_{\mu\nu}$
to a fixed point (modulo at most a finite number of relevant
parameters).  The fixed points are the solutions of
\eq
\label{eq:fixedpoint}
R_{\mu\nu} = \nabla_{\mu}v_{\nu}+\nabla_{\mu}v_{\mu}
\en
where $v^{\mu}(x)\partial_{\mu}$ is a vector field on $M$.  A vector field
is an infinitesimal reparametrization $\dot x^{\mu}=
v^{\mu}(x)$ of $M$.  This is a field redefinition in the nonlinear model
equivalent to the redundant perturbation $\dot g_{\mu\nu}
=\nabla_{\mu}v_{\nu}+\nabla_{\nu}v_{\mu}$.  Thus the fixed point equation
(\ref{eq:fixedpoint}) expresses physical 2-d scale invariance\footnote{\label{fn1}
The general nonlinear model (also called the {\it nonlinear sigma
model}) was constructed in
\cite{Friedan1980a,Friedan1980c,Friedan:1980jmcosmo}
%\cite{Friedan1980a,Friedan1980c,Friedan:1980jm}
in $2{+}\epsilon$ dimensions with fixed point 
equation $R_{\mu\nu} -\epsilon g_{\mu\nu} = 
\nabla_{\mu}v_{\nu}+\nabla_{\mu}v_{\mu}$.   The solutions were 
called {quasi-Einstein metrics}.  The Ricci flow was 
introduced in Mathematics in \cite{hamilton1982}.
The solutions of $R_{\mu\nu} -\epsilon g_{\mu\nu} = 
\nabla_{\mu}v_{\nu}+\nabla_{\mu}v_{\mu}$
have been called \emph{Ricci solitons} or \emph{Ricci flow solitons} in the Mathematics literature.}.

The 2-d quantum field theory is manifestly well defined when $g_{\mu\nu}$ 
has euclidean signature.  Assume that $M$ is a
four-dimensional Riemannian manifold of the form $I\times S^{3}$ where $I$ is a
real interval and $S^{3}$ is the unit 3-sphere.  Assume $\SO(4)$
invariance.  Parametrize $I\times S^{3}$ as a spherical shell in
$\Reals^{4}$.  The flat euclidean metric is
$\delta_{\mu\nu}dx^{\mu}dx^{\nu}= dr^{2} +r^{2} ds^{2}_{S^{3}}$ where
$ds^{2}_{S^{3}}$ is the round metric on the unit 3-sphere.  The general
$\SO(4)$-invariant metric has the form
$g_{\mu\nu}dx^{\mu}dx^{\nu}=F_{1}(r)^{2}dr^{2} +F_{2}(r)^{2}
ds^{2}_{S^{3}}$.  The metric can be made conformally flat
\eq
g_{\mu\nu}dx^{\mu}dx^{\nu} =  a^{2} \left( d\tau^{2} + 
ds^{2}_{S^{3}} \right)
\qquad
a = e^{f(\tau)}
\en
by reparametrizing $r\rightarrow \tau(r)$ with $dr/d\tau = 
F_{2}/F_{1}$.
The general $\SO(4)$-invariant vector field is 
$v^{\tau}(\tau)\partial_{\tau}$.
Then \cite{cosmonote}\nocite{sagemath:2019:8.8}
%\footnote{
% See Supplemental Material at [URL will be inserted by publisher]
%
%Calculations are shown in the attached supplements \cite{cosmonote}.
% Numerical calculations are done in
% SageMath notebooks \cite{sagemath:2019:8.7}.}
% The Sage Developers, {\it SageMath, the Sage Mathematics Software System (Version 8.7)} (2019),
% \url{https://www.sagemath.org}.}
%\cite{sagemath:2019:8.7}.}
% and TIDES \cite{Abad:2015:AIN:2831512.2831799}
%are carried out in a Jupyter notebook.}
\eq
\begin{gathered}
\begin{aligned}
R_{\mu\nu}dx^{\mu}dx^{\nu} &=
\left(-3\partial_{\tau}f_{\tau}\right) d\tau^{2}
+
\left(-\partial_{\tau}f_{\tau}+2 - 2f_{\tau}^{2}\right)
ds^{2}_{S^{3}}
\\[1ex]
(\nabla_{\mu}v_{\nu}+\nabla_{\nu}v_{\mu})dx^{\mu}dx^{\nu}
&=
\left(2\partial_{\tau}v_{\tau}-2v_{\tau}f_{\tau} \right) d\tau^{2}
+
\left( 2v_{\tau}f_{\tau}\right) ds^{2}_{S^{3}}
\end{aligned}
\\[1ex]
f_{\tau}=\partial_{\tau}f
\qquad
v_{\tau} = g_{\tau\tau}v^{\tau} = a^{2} v^{\tau}
\end{gathered}
\en
The fixed point equation (\ref{eq:fixedpoint}) becomes
the ordinary differential equation
\eq
\label{eq:ode}
\partial_{\tau}f_{\tau} = -2f_{\tau}v_{\tau} -2f_{\tau}^{2}+2
\qquad
\partial_{\tau}v_{\tau} = 4 f_{\tau}v_{\tau} +3 f_{\tau}^{2}-3
\en
We shall see below that there is an essentially
unique solution of
(\ref{eq:ode}) which analytically continues in $\tau$ to real time $T
= i^{-1}\tau$
\eq
ds^{2} = a^{2} \left(-dT^{2} + ds^{2}_{S^{3}}\right)
\en
The real-time ode is
\eq
\label{eq:odeT}
\begin{gathered}
\partial_{T}f_{T} = -2f_{T}v_{T} -2f_{T}^{2}-2
\qquad
\partial_{T}v_{T} = 4 f_{T}v_{T} +3 f_{T}^{2}+3
\\[1ex]
\partial_{T} = i\partial_{\tau}
\qquad
f_{T}= i f_{\tau} = \partial_{T}a/a
\qquad
v_{T} = i v_{\tau}
\end{gathered}
\en
The solution is
\eq
\label{eq:cosmosolution}
f_{T}=\frac{\cos 2T +\sqrt3}{\sin 2T}
\qquad
v_{T} = \frac{-\sqrt3}{\sin 2T}
\qquad
a =t'_{0} \sin^{1+\nu} T \cos^{-\nu} T
\qquad
T \in (0, \pi/2)
\en
with $\nu = \sqrt3/2-1/2= 0.3660\ldots$~.
The only free parameter is the overall time scale $t'_{0}$.
In co-moving time $t$
\eq
\label{eq:comovingt}
\begin{gathered}
ds^{2} = -dt^{2} + a^{2} ds^{2}_{S^{3}}
\qquad
dt = a dT
\\[1ex]
\frac{t}{t'_{0}}=
\int_{0}^{T} \sin^{1+\nu} T' \cos^{-\nu} T'\, dT'
=\frac12 B_{\sin^{2}T}\left(\frac{2+\nu}{2},\frac{1-\nu}{2} \right)
\\[1ex]
t\in (0,t_{\max})
\qquad
\frac{t_{\max}}{t'_{0}}=  \frac12 B\left(\frac{2+\nu}{2},\frac{1-\nu}{2} \right)
= 1.470\ldots
\end{gathered}
\en
$B(p,q)$ is the Euler beta function,
$B_{x}(p,q)$ the incomplete beta function.
At the limits
\eq
\begin{alignedat}{3}
T&\rightarrow 0
\qquad&
\frac{t}{t_{0}'} &\rightarrow \frac {T^{2+\nu}}{2+\nu}
\qquad&
a &\rightarrow t_{0}' 
(2+\nu)^{1/\sqrt3} 
\left(
\frac{t}{t_{0}'}
\right)^{1/\sqrt3}
\\
T&\rightarrow \frac{\pi}{2}
\qquad&
\frac{t_{\max}-t}{t_{0}'} &\rightarrow \frac 
{\left(\frac{\pi}{2}-T\right)^{1-\nu}}{1-\nu}
\qquad&
a &\rightarrow  t_{0}' 
(1-\nu) ^{-1/\sqrt3}
\left(
\frac{t_{\max}-t}{t_{0}'}
\right)^{-1/\sqrt3}
\end{alignedat}
\en
This is an expanding universe which begins with a big bang 
$a \sim t^{0.577\ldots}$
and ends with a big blowup  $a \sim 
(t_{\max}-t)^{-0.577\ldots}$.
The Hubble parameter $H= \partial_{t}a/a = f_{T}/a$ is
\eq
\label{eq:H}
H =  \frac{1}{t'_{0}}
(\cos^{2}T +\nu)
\sin^{-2-\nu} T \cos^{-1+\nu} T
\en
The deceleration parameter
$q  = -a\partial_{t}^{2}a/(\partial_{t}a)^{2} 
= -\partial_{T}f_{T}/f_{T}^{2}$ is
\eq
\label{eq:q}
q 
=\frac{2\left( \sqrt3\cos 2T+1\right)}{\left(\cos2T  +\sqrt3\right)^{2}}
\qquad
T_{q{=}0}=
\frac12 \arccos(-1/\sqrt3)
= (0.6959\ldots) \frac{\pi}{2}
\en
The expansion decelerates ($q>0$) until $T=T_{q{=}0}$ 
then accelerates ($q<0$)
until the end.

For a first estimate of the time scale $t_{0}'$ and the present time 
$t_{0}$
use
the deceleration parameter at the present time
$q_{0}=q(T_{0})\approx -0.6$ 
in equation (\ref{eq:q}) to obtain $T_{0}=0.77 \,\pi/2$.
Then equate
the Hubble constant $H_{0}$ %\approx  68\, \mathrm{km/s/Mpc}
to $H(T_{0})$ in (\ref{eq:H}) to obtain $t_{0}'= 1.1\, t_{H} $
where $t_{H}=1/H_{0} \approx 
1.4\times 10^{10}y$.
Then (\ref{eq:comovingt}) gives $t_{\max} = 1.6\, t_{H}$, $t_{0}= 
0.73\, t_{H}$
and (\ref{eq:cosmosolution}) gives $a_{0}=a(t_{0}) = 
1.5\, t_{H}$.

\begin{figure}
\includegraphics[scale=1]{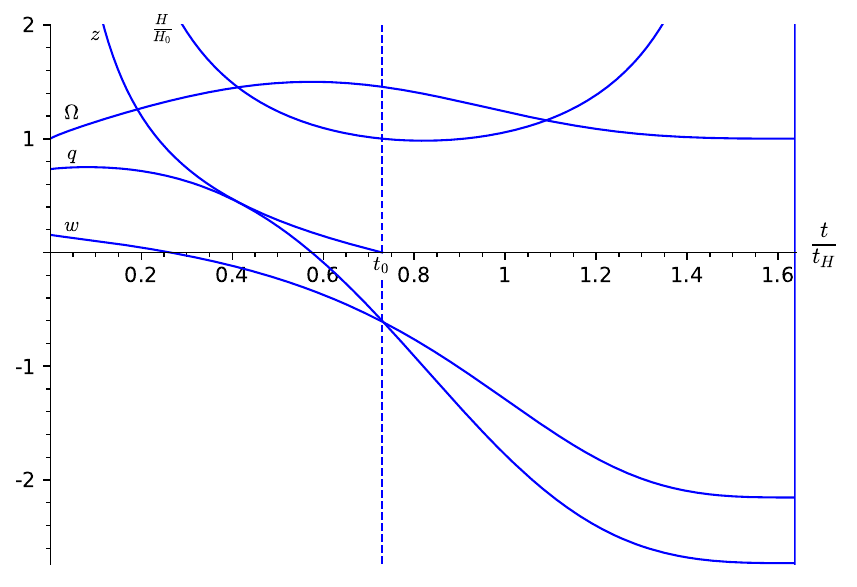}
\caption{The cosmological parameters
as functions of the co-moving time $t$.
The present is $t_{0}$.
\label{fig:params}}
\end{figure}

Einstein's equation
\eq
R_{\mu\nu} -\frac12 R g_{\mu\nu} = 8\pi G \,T_{\mu\nu}
\en
is satisfied with energy-momentum tensor
\eq
T_{\mu\nu} = \frac1{8\pi G} \left(
\nabla_{\mu}v_{\nu}+\nabla_{\nu}v_{\mu}
- \nabla_{\sigma}v^{\sigma} g_{\mu\nu}
\right)
\en
which is that of a perfect fluid
of density $\rho(t)$ and pressure $p(t)$
\eq
\begin{gathered}
T_{\mu\nu}dx^{\mu}dx^{\nu}
= \rho(t) dt^{2} + p(t) a^{2} ds^{2}_{S^{3}}
\\[1ex]
8\pi G \,\rho(t) = a^{-2}\left(3f_{T}^{2}+3
\right)
\qquad
8\pi G \,p(t) = a^{-2}\left(4 f_{T}v_{T} 
+3 f_{T}^{2}
+3 
\right)
\end{gathered}
\en
The energy-momentum is purely gravitational, tautologically dark.
The equation-of-state parameter $w(t) = p/\rho$ 
and
the density parameter $\Omega(t) = 8\pi G\rho/3 H^{2}=1+1/f_{T}^{2}$ 
are
\eq
w =  \frac{\cos 2T}{3\cos 2T + 2\sqrt3}
\qquad
\Omega = 1+ \left( \frac{\sin 2T}{\cos 2T + \sqrt3}\right)^{2}
\en
The estimate $T_{0}=0.77\,\pi/2$ gives present values
$w_{0}=-0.6$, $\Omega_{0} =1.5$.  
The cosmological parameters are graphed as functions of $t$ in
Figure~\ref{fig:params}.
Past values for a selection of redshifts $z = a_{0}/a -1$ are listed
in Table~\ref{table:retrodictions}.  
%Figure~\ref{fig:conf} shows the conformal diagram.
%

%%%%%%%%%%%%%%%%%%%%%%%%%%%%%%%%%%%%%%%%%%%%%%%%%%%%%%%%%%%%%%%%
%                                                              %
%   \begin{figure}                                             %
%   \includegraphics[scale=1]{figures/conformal_diagram.pdf}   %
%   \caption{The conformal diagram.                            %
%   $R$ is geodesic distance on the unit 3-sphere.             %
%   $T$ is conformal time.                                     %
%   $T_{0}$ is the present.                                    %
%   \label{fig:conf}}                                          %
%   \end{figure}                                               %
%                                                              %
%%%%%%%%%%%%%%%%%%%%%%%%%%%%%%%%%%%%%%%%%%%%%%%%%%%%%%%%%%%%%%%%
%
\begin{table}
\caption{\label{table:retrodictions}Cosmological parameters at 
selected values of $z$}
\begin{ruledtabular}
\begin{tabular}{c|rcccc}
 & $z$ & $H/H_0$ & $q$ & $w$ & $\Omega$ \\ \hline
\rule{0pt}{3ex}
$z\gg 1$ & $\gg 1$ & 0.75 $z^{1.7}$ & $0.73$ & $0.16$ & $1.0$ \\
 & 1000 & $1.2\times 10^{5}$ & $0.73$ & $0.16$ & $1.0$ \\
 & 100 & $2.2\times 10^{3}$ & $0.73$ & $0.15$ & $1.0$ \\
 & 10 & $48$ & $0.74$ & $0.15$ & $1.0$ \\
$T=\frac12 T_0$ & 1.7 & $4.1$ & $0.74$ & $0.08$ & $1.2$ \\
 & 1.0 & $2.4$ & $0.69$ & $0.02$ & $1.3$ \\
$q=0$ & 0.2 & $1.1$ & $0$ & $-0.3$ & $1.5$ \\
$t=t_0$ & 0 & $1$ & $-0.6$ & $-0.6$ & $1.5$ \\
\end{tabular}
\end{ruledtabular}
\end{table}

We analyze the real-time ode (\ref{eq:odeT}) by adapting some of the methods that were used
in \cite{Bryant2}
to analyze the ode analogous to (\ref{eq:ode}) for the euclidean-signature 
$\SO(3)$-invariant
fixed point equation in three dimensions.
The real-time ode (\ref{eq:odeT}) has time-reflection symmetry 
$(T,f_{T},v_{T})\rightarrow (-T,-f_{T},-v_{T})$
and constant of motion
\eq
C= \frac1{a^{2}}\left(\frac23 v_{T}^{2}+2f_{T}v_{T} +  f_{T}^{2} + 1\right)
\qquad
\partial_{T} C = 0
\en
Changing variables to $h_{\pm} = f_{T}+ (1\pm1/\sqrt3) v_{T}$
the constant of motion and ode become
\eq
\label{eq:odehpm}
h_{+}h_{-}+1 = C a^{2}
\qquad
\partial_{T}h_{\pm} = -(1 + h_{\pm}^{2}) + \lambda_{\pm}(h_{+}h_{-}+1)
\qquad
\lambda_{\pm} = 2\pm\sqrt3
\en
The cosmological solution (\ref{eq:cosmosolution}) is
$C=0$, $h_{-}=\cot T$, $h_{+}=- \tan T$.

\begin{figure}
\subfloat{
  \includegraphics[scale=0.7]{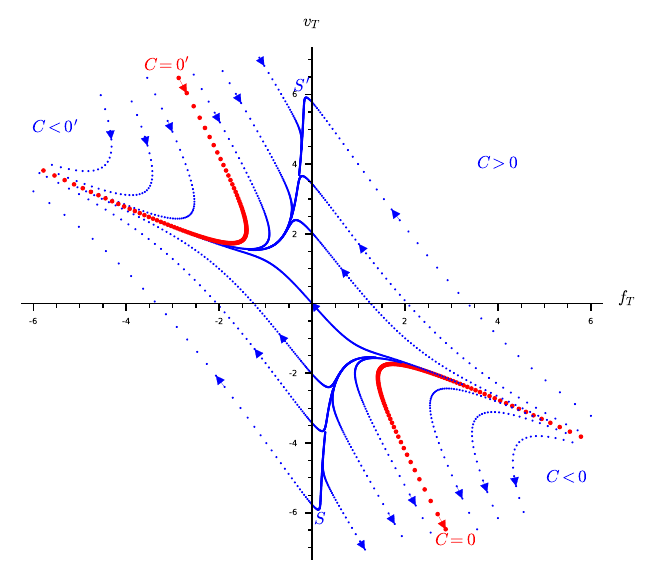}
  }
\subfloat{
  \includegraphics[scale=0.7]{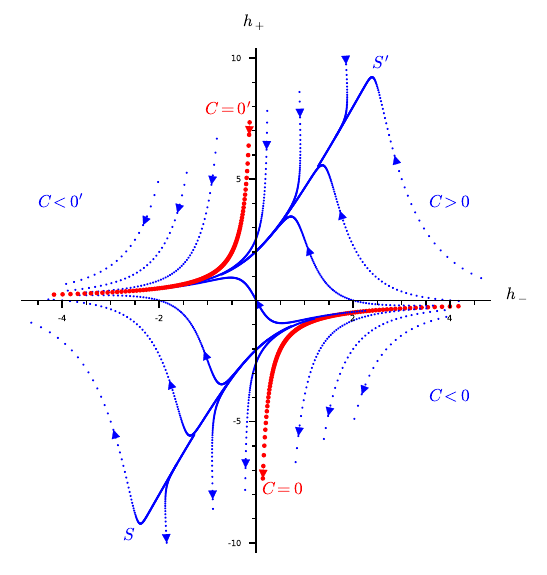}
  }
\caption{Phase portrait of the real-time ode (\ref{eq:odeT},\ref{eq:odehpm})
in terms of $(f_{T},v_{T})$ on the left,
$(h_{-},h_{+})$ on the right.  The trajectories are the solutions.  The 
arrows point in the direction of increasing $T$.  The interval 
between points in the trajectories is $\Delta T = 0.01$.  The universe is 
expanding when $f_{T}=\partial_{t}a$ is positive.  The expansion is accelerating when $f_{T}$ 
is increasing.
The two red trajectories are the $C=0$ solutions.  The one in the 
lower right quadrant labeled $C{=}0$ is the cosmological solution~(\ref{eq:cosmosolution}).
Its time reflection is labeled $C{=}0'$.
The separatrix is labeled ${S}$. Its time-reflection 
is labeled ${S}'$.
\label{fig:phaseportrait}}
\end{figure}

The phase portrait of the ode
is shown in 
Figure~\ref{fig:phaseportrait}.
Every $C\ne 0$ trajectory asymptotes to a $C=0$ trajectory.
The separatrix ${S}$ and its time-reflection ${S'}$
each does so at one end; all the 
other $C\ne 0$ trajectories do so at both ends.
The asymptotic behavior is one of
\eq
T \rightarrow 0^{\pm}
\qquad
\begin{alignedat}{3}
h_{-}&\rightarrow \frac{1}{T}
\qquad &
h_{+} &\rightarrow -T - \frac13 T^{3} + c T (T^{2})^{1+\nu} 
\qquad&
\frac{c}{C} &> 0
\\[1ex]
h_{+}&\rightarrow \frac{1}{T}
\qquad &
h_{-} &\rightarrow   c'  T (T^{2})^{-\nu}
\qquad&
\frac{c'}{C} &> 0
\end{alignedat}
\en
The presence of an irrational power of $T$,
namely $(T^{2})^{1+\nu}$ or $ (T^{2})^{-\nu}$,
implies that
such a solution cannot be the analytic continuation
of a solution in imaginary time $\tau = i T$.

The trajectories that start from $h_{-}=\infty$, $h_{+}=0$ 
at $T=0$ are 
characterized by the numerical invariant
\eq
c 
=
(h_{+}h_{-}+1)
(h_{-}^{2})^{1+\nu}
e^{\int_{0}^{T} \frac{2\,dT'}{h_{-}(T')} }
= \lim_{T\rightarrow 0} (h_{+}h_{-}+1) (h_{-}^{2})^{1+\nu}
\en
The separatrix $S$ has $c_{S}=0.284\ldots$~.
The $c<c_{S}$ trajectories --- the trajectories below $S$ ---
are all roughly realistic as cosmologies.
They all start with a big bang $a\sim 
t^{1/\sqrt3}$ then decelerate than accelerate to a big blow-up
$a\sim t^{-1/\sqrt3}$, all driven by purely 
gravitational dark energy-momentum.
The gap between the $C{=}0$ cosmological solution and the separatrix ${S}$
means that the cosmological solution~(\ref{eq:cosmosolution})
is stable against real-time perturbations.  The zone of stability
on the side $C>0$
shrinks to zero as $T\rightarrow 0$.

The two-dimensional renormalization group was developed into a
comprehensive fundamental theory of physics in which the 
2-d renormalization group acts not on a space of classical space-time fields but rather 
on measures on that space, producing not a solution of the classical 
field equations but rather a measure --- a functional integral ---
expressing a solution of the quantum field equations 
\cite{Friedan2003a,Friedan2010ua,Friedan:2018rnd}.
The present exploratory calculations of cosmology from the 2-d 
renormalization group will
need to be connected to this fundamental theory.
Assumptions such as $SO(4)$ space-time symmetry
will need to be justified by 
properties of the 2-d renormalization group flow --- by stability properties of the 
fixed point and/or as the dynamical result of a distinguished 
trajectory of the rg flow.

The analyticity assumption needs precise characterization and
justification.
The assumption here that $f_{T}(T)$ and $v_{T}(T)$ should be analytic 
in $T$
singles out the $C=0$ cosmological solution.
Analyticity serves as a selection principle
providing specificity.
It should express a physical principle.

The next exploratory steps
will add standard model fields
to the calculation,
then spatial fluctuations
and quantum effects,
hoping to find
analytic solutions of
2-d renormalization group fixed point 
equations
that capture more and more qualitative features of the cosmology of the real 
world,
aiming to
find a solution
that can be tested in quantitative detail.
This would provide support for the fundamental theory \cite{Friedan2003a} in 
which a quantum version of the two-dimensional renormalization group 
acts mechanically to produce physics
and would provide guidance for deriving
real world cosmology from that fundamental theory.

\begin{acknowledgments}
This work was supported by the Rutgers New High Energy Theory Center
and by the generosity of B. Weeks.
I am grateful
to the  Mathematics Division of the 
Science Institute of the University of Iceland
for its hospitality.
\end{acknowledgments}

\vspace*{4ex}

\bibliography{Literature}

\begin{thebibliography}{10}%
\makeatletter
\providecommand \@ifxundefined [1]{%
 \@ifx{#1\undefined}
}%
\providecommand \@ifnum [1]{%
 \ifnum #1\expandafter \@firstoftwo
 \else \expandafter \@secondoftwo
 \fi
}%
\providecommand \@ifx [1]{%
 \ifx #1\expandafter \@firstoftwo
 \else \expandafter \@secondoftwo
 \fi
}%
\providecommand \natexlab [1]{#1}%
\providecommand \enquote  [1]{``#1''}%
\providecommand \bibnamefont  [1]{#1}%
\providecommand \bibfnamefont [1]{#1}%
\providecommand \citenamefont [1]{#1}%
\providecommand \href@noop [0]{\@secondoftwo}%
\providecommand \href [0]{\begingroup \@sanitize@url \@href}%
\providecommand \@href[1]{\@@startlink{#1}\@@href}%
\providecommand \@@href[1]{\endgroup#1\@@endlink}%
\providecommand \@sanitize@url [0]{\catcode `\\12\catcode `\$12\catcode
  `\&12\catcode `\#12\catcode `\^12\catcode `\_12\catcode `\%12\relax}%
\providecommand \@@startlink[1]{}%
\providecommand \@@endlink[0]{}%
\providecommand \url  [0]{\begingroup\@sanitize@url \@url }%
\providecommand \@url [1]{\endgroup\@href {#1}{\urlprefix }}%
\providecommand \urlprefix  [0]{URL }%
\providecommand \Eprint [0]{\href }%
\providecommand \doibase [0]{https://doi.org/}%
\providecommand \selectlanguage [0]{\@gobble}%
\providecommand \bibinfo  [0]{\@secondoftwo}%
\providecommand \bibfield  [0]{\@secondoftwo}%
\providecommand \translation [1]{[#1]}%
\providecommand \BibitemOpen [0]{}%
\providecommand \bibitemStop [0]{}%
\providecommand \bibitemNoStop [0]{.\EOS\space}%
\providecommand \EOS [0]{\spacefactor3000\relax}%
\providecommand \BibitemShut  [1]{\csname bibitem#1\endcsname}%
\let\auto@bib@innerbib\@empty
%</preamble>
\bibitem [{\citenamefont {Friedan}(1980)}]{Friedan1980a}%
  \BibitemOpen
  \bibfield  {author} {\bibinfo {author} {\bibfnamefont {D.}~\bibnamefont
  {Friedan}},\ }\bibfield  {title} {\bibinfo {title} {{Nonlinear Models in
  $2+\epsilon$ Dimensions}},\ }\href
  {https://doi.org/10.1103/PhysRevLett.45.1057} {\bibfield  {journal} {\bibinfo
   {journal} {Phys. Rev. Lett.}\ }\textbf {\bibinfo {volume} {45}},\ \bibinfo
  {pages} {1057} (\bibinfo {year} {1980})}\BibitemShut {NoStop}%
%%CITATION = PRLTA,45,1057;%%
\bibitem [{\citenamefont {{Daniel Harry Friedan}}(1980)}]{Friedan1980c}%
  \BibitemOpen
  \bibfield  {author} {\bibinfo {author} {\bibnamefont {{Daniel Harry
  Friedan}}},\ }\href@noop {} {\emph {\bibinfo {title} {{Nonlinear Models in
  $2+\epsilon$ Dimensions}}}},\ \bibinfo {type} {Tech. Rep.}\ (\bibinfo
  {institution} {Lawrence Berkeley Laboratory LBL-11517},\ \bibinfo {year}
  {1980})\ \bibinfo {note} {{U.C. Berkeley PhD thesis (1980)}}\BibitemShut
  {NoStop}%
\bibitem [{\citenamefont {{Daniel Harry Friedan}}(1985)}]{Friedan:1980jmcosmo}%
  \BibitemOpen
  \bibfield  {author} {\bibinfo {author} {\bibnamefont {{Daniel Harry
  Friedan}}},\ }\bibfield  {title} {\bibinfo {title} {{Nonlinear Models in
  $2+\epsilon$ Dimensions}},\ }\href
  {https://doi.org/10.1016/0003-4916(85)90384-7} {\bibfield  {journal}
  {\bibinfo  {journal} {Ann. Phys.}\ }\textbf {\bibinfo {volume} {163}},\
  \bibinfo {pages} {318} (\bibinfo {year} {1985})},\ \bibinfo {note}
  {{republication of [2]}}\BibitemShut {NoStop}%
%%CITATION = APNYA,163,318;%%
\bibitem [{\citenamefont {{Richard S. Hamilton}}(1982)}]{hamilton1982}%
  \BibitemOpen
  \bibfield  {author} {\bibinfo {author} {\bibnamefont {{Richard S.
  Hamilton}}},\ }\bibfield  {title} {\bibinfo {title} {{Three-manifolds with
  positive Ricci curvature}},\ }\href {https://doi.org/10.4310/jdg/1214436922}
  {\bibfield  {journal} {\bibinfo  {journal} {J. Differential Geom.}\ }\textbf
  {\bibinfo {volume} {17}},\ \bibinfo {pages} {255} (\bibinfo {year}
  {1982})}\BibitemShut {NoStop}%
\bibitem [{cos()}]{cosmonote}%
  \BibitemOpen
  \href@noop {} {}\bibinfo {howpublished} {Calculations are shown in the
  accompanying supplemental material which consists of a note {\it Calculations
  for "Cosmology \ldots "} and three SageMath \cite{sagemath:2019:8.8}
  notebooks of numerical calculations. The supplemental material can be found
  at
  \url{http://www.physics.rutgers.edu/pages/friedan/papers/2019/Cosmology_I_supplementary_material/}
  or at
  \url{https://share.cocalc.com/share/6a7035ba-9879-4b05-b3d0-688b1309a21c/Cosmology_I_supplementary_material/?viewer=share/}.}\BibitemShut
  {Stop}%
\bibitem [{\citenamefont {{The Sage Developers}}(2019)}]{sagemath:2019:8.8}%
  \BibitemOpen
  \bibfield  {author} {\bibinfo {author} {\bibnamefont {{The Sage
  Developers}}},\ }\href@noop {} {\emph {\bibinfo {title} {{S}ageMath, the
  {S}age {M}athematics {S}oftware {S}ystem ({V}ersion 8.8)}}} (\bibinfo {year}
  {2019}),\ \bibinfo {note} {{\tt https://www.sagemath.org}}\BibitemShut
  {NoStop}%
\bibitem [{\citenamefont {{{Robert L. Bryant}}}(2005)}]{Bryant2}%
  \BibitemOpen
  \bibfield  {author} {\bibinfo {author} {\bibnamefont {{{Robert L.
  Bryant}}}},\ }\href {https://math.duke.edu/~bryant/3DRotSymRicciSolitons.pdf}
  {\bibinfo {title} {{{Ricci flow solitions in dimension 3 with
  SO(3)-symmetries}}}},\ \bibinfo {howpublished} {unpublished} (\bibinfo {year}
  {2005}),\ \bibinfo {note}
  {{{\url{https://math.duke.edu/~bryant/3DRotSymRicciSolitons.pdf}}}}\BibitemShut
  {NoStop}%
\bibitem [{\citenamefont {Friedan}(2003)}]{Friedan2003a}%
  \BibitemOpen
  \bibfield  {author} {\bibinfo {author} {\bibfnamefont {D.}~\bibnamefont
  {Friedan}},\ }\bibfield  {title} {\bibinfo {title} {{A tentative theory of
  large distance physics}},\ }\href@noop {} {\bibfield  {journal} {\bibinfo
  {journal} {JHEP}\ }\textbf {\bibinfo {volume} {10}},\ \bibinfo {pages}
  {063}},\ \Eprint {https://arxiv.org/abs/hep-th/0204131}
  {arXiv:hep-th/0204131} \BibitemShut {NoStop}%
%%CITATION = HEP-TH/0204131;%%
\bibitem [{\citenamefont {Friedan}(2010)}]{Friedan2010ua}%
  \BibitemOpen
  \bibfield  {author} {\bibinfo {author} {\bibfnamefont {D.}~\bibnamefont
  {Friedan}},\ }\bibfield  {title} {\bibinfo {title} {{A loop of $SU(2)$ gauge
  fields stable under the Yang-Mills flow}},\ }in\ \href@noop {} {\emph
  {\bibinfo {booktitle} {{Perspectives in mathematics and physics: Essays
  dedicated to Isadore Singer's 85th birthday}}}},\ \bibinfo {series} {Surveys
  in Differential Geometry}, Vol.~\bibinfo {volume} {15},\ \bibinfo {editor}
  {edited by\ \bibinfo {editor} {\bibfnamefont {T.}~\bibnamefont {Mrowka}}\
  and\ \bibinfo {editor} {\bibfnamefont {S.-T.}\ \bibnamefont {Yau}}}\
  (\bibinfo  {publisher} {International Press of Boston},\ \bibinfo {year}
  {2010})\ pp.\ \bibinfo {pages} {131--204},\ \Eprint
  {https://arxiv.org/abs/1008.1189} {arXiv:1008.1189 [hep-th]} \BibitemShut
  {NoStop}%
%%CITATION = 1008.1189;%%
\bibitem [{\citenamefont {Friedan}(2018)}]{Friedan:2018rnd}%
  \BibitemOpen
  \bibfield  {author} {\bibinfo {author} {\bibfnamefont {D.}~\bibnamefont
  {Friedan}},\ }\bibfield  {title} {\bibinfo {title} {A pragmatic approach to
  formal fundamental physics},\ }\href@noop {} {\  (\bibinfo {year} {2018})},\
  \Eprint {https://arxiv.org/abs/1810.09508} {arXiv:1810.09508 [hep-th]}
  \BibitemShut {NoStop}%
%%CITATION = ARXIV:1810.09508;%%
\end{thebibliography}%

\end{document}